# Characterization of Sine- Skewed von Mises Distribution

Mohammad Ahsanullah  
Department of Management Sciences  
Rider University  
Lawrenceville, New Jersey, USA

M. Z. Anis[1]  
SQC & OR Unit  
Indian Statistical Institute  
203, B T Road  
Calcutta 700 108, India

**Abstract**

The von Mises distribution is one of the most important distribution in statistics to deal with circular data. In this paper we will consider some basic properties and characterizations of the sine skewed von Mises distribution.

**Key words**- Characterizations, Sine Skewed von Mises Distribution, Truncated First Moment.

**2010 Mathematics Subject Classifications**: 62E10, 62E15, 62G30

**1. Introduction**

Skew-symmetric distributions were introduced by Azzalini (1985). It was presented as a skewing normal distribution. This technique is used to skew any continuous symmetric distribution. See - for example - in the univariate case, Arnold and Beaver (2000), Azzalini (2014), Gupta et al. (2002) and Umbach and Jammalamadaka (2009). Abe and Pewsey (2011) considered sine-skewed circular distribution.

Suppose $f_\theta(\theta)$ is a symmetric circular density of a circular random variable $\theta$ which is symmetric about zero, then sine-skewed circular probability density $f_\theta(\theta_1)$ is defined as
$$f_\theta(\theta_1) = f_0(\theta_1)(1 + \lambda \sin \theta_1) \qquad (1.1)$$
where $-\pi \leq \theta_1 \leq \pi$ and $-1 \leq \lambda \leq 1$. It should be noted that our construction of the sine-skewed circular probability density is slightly different from that of Abe and Pewsey (2011). In this paper $f_0(\theta_1)$ will be taken as von Mises distribution with $f_0(\theta_1)$ as
$$f_0(\theta_1) = \frac{e^{k \cos \theta_1}}{2\pi I_0(k)}, \quad -\pi \leq \theta_1 \leq \pi, \quad k > 0 \qquad (1.2)$$
where $\theta_1$ is measured in radian, $k$ is a scale factor and $I_0(k)$ is the modified Bessel function of the first kind and order 0.

It is known (see Abramowitz and Stegan (1970)) that
$$I_0(k) = \sum_{j=0}^{\infty} \left(\frac{k}{2}\right)^2 \left(\frac{1}{j!}\right)^2.$$

---

[1] Corresponding Author

The von Mises distribution is one of the most important distributions in statistics to deal with the circular data or directional data. It is the circular analog of the normal distribution on a line. Combining (1.1) and (1.2), the sine-skewed von Mises circular distribution with probability density function (pdf) $f_{svm}(\theta_1)$ will be

$$f_{svm}(\theta_1) = \frac{e^{k\cos\theta_1}}{2\pi I_0(k)}(1 + \lambda \sin\theta_1), \quad -\pi \leq \theta_1 \leq \pi; \ -1 \leq \lambda \leq 1 \text{ and } k \geq 0. \quad (1.3)$$

We will denote the random variable $\theta$ with pdf as given in (1.3) above as $vM(\theta, \lambda, k)$.

While inferential issues with respect to this distribution has been discussed in the literature, see for example, Ley and Verdebout (2014) and Abe and Ley (2017), to the best of our knowledge no characterization results are available. We aim to fill this gap. The paper is organized as follows. Section 2 gives the main results; while Section 3 presents two characterizations of the skew von Mises distribution. Finally Section 4 concludes the paper.

**2. Main Results**

It is known that

$$e^{k\cos\theta} = I_0(k) + 2\sum_{j=1}^{\infty} I_j(k)\cos j\theta, \quad (2.1)$$

where $I_j(k)$ is the Bessel function of order $j$.
Using (2.1) the probability density function (pdf) of the skew von Mises distribution can be written as

$$f_{svm}(\theta_1, \lambda, k) = \frac{1}{2\pi I_0(k)}\left(I_0(k) + 2\sum_{j=1}^{\infty} I_j(k)\cos j\theta_1\right)(1 + \lambda \sin\theta_1), \quad (2.2)$$

where $-\pi \leq \theta_1 \leq \pi; \ -1 \leq \lambda \leq 1$ and $k \geq 0$.

The Figure 2.1. gives the pdf of $f(\theta, \lambda, k)$ for $k = 1$ and $\lambda == -0.2; -0.5; 0.2$ and $0.5$.

The cumulative distribution function (cdf) of skew von Mises distribution with the probability density function (pdf) as given in (1.3) can be written as

$$F_{svm}(\theta_1) = \frac{1}{2\pi}\left\{(\pi + \theta_1) + 2\sum_{j=1}^{\infty}\frac{I_j(k)}{j}\sin(j\theta_1)\right\} + \frac{\lambda}{2\pi I_0(k)}\left(e^{-k} - e^{k\cos\theta_1}\right) \quad (2.3)$$

where

$$I_j(k) = \left(\frac{k}{2}\right)^j \sum_{i=0}^{\infty}\left\{\left(\frac{k}{2}\right)^{2i}\left(\frac{1}{i!\,(j+1)!}\right)\right\}, \quad -\pi \leq \theta_1 \leq \pi, \quad -1 \leq \lambda \leq 1, \text{ and } k \geq 0.$$

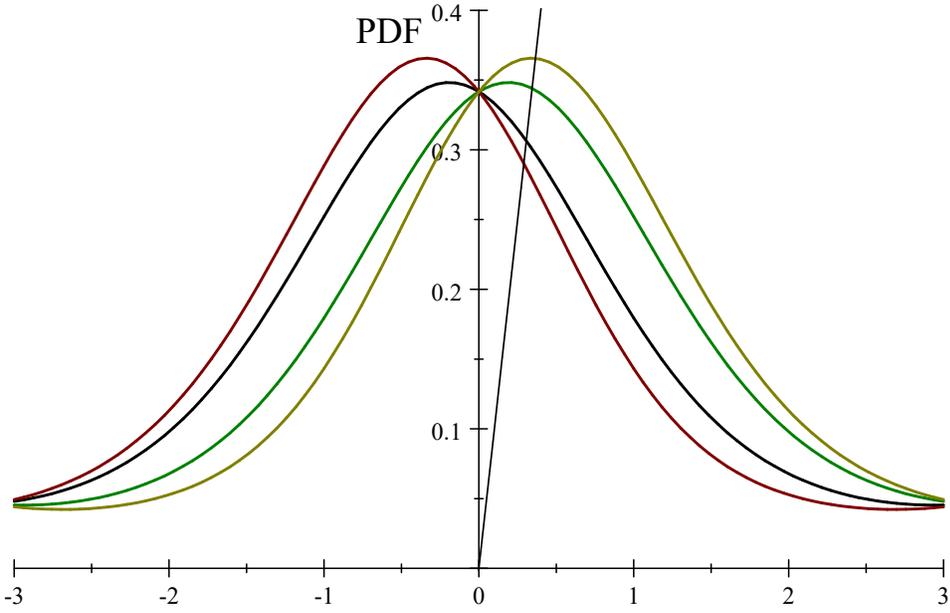

Fig 2.1. Pdfs $f(\theta, -0.2, 1)$ – Black; $f(\theta, -0.4, 1)$ – Red; $f(\theta, 0.2, 1)$ – Green; and $f(\theta, 0.4, 1)$ – Brown.

The cdf $F(\theta, \lambda_1)$ is shown in Figure 2.2.

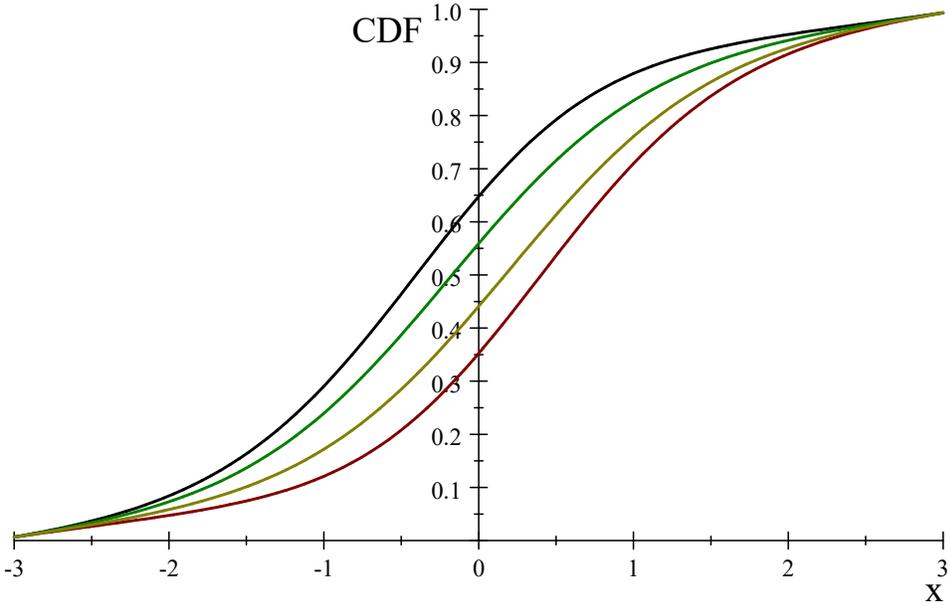

Fig.2.2. Cdf's $F_{svm}(\theta, -0.5, 1)$- black; $F_{svm}(\theta, 0.5, 1)$- red; $F_{svm}(\theta, -0.2, 1)$- green; $F_{svm}(\theta, 0.2, 1)$- brown.

We have $k = \frac{1}{2} \ln \frac{f(0)}{f(\pi)}$.

The mode is obtained by solving the equation $\frac{d}{d\theta} f_{svm}(\theta, \lambda, k) = 0$.

Now,
$$\frac{d}{d\theta}\left(\frac{e^{k\cos\theta}}{2\pi I_0(k)}(1+\lambda\sin\theta)\right) = 0$$

implies
$$k\lambda(\sin\theta)^2 - \lambda\cos\theta + k\sin\theta = 0 \tag{2.4}$$

Thus, the mode is a solution to the equation in (2.4) above. Furthermore, if for any fixed $k$ the mode of $F_{svm}(\theta, \lambda, k)$ is at $\theta_0$, then the mode of $F_{svm}(\theta, -\lambda, k)$ will be at $-\theta_0$. The following Table gives the mode of $F_{svm}(\theta, \lambda, k)$ for $k = 1, 2$ and $10$, and $\lambda = 0.1$ to $0.9$.

Table 2.1. Mode of $vM(\theta, \lambda, k)$

| $k \downarrow \;\; \lambda \rightarrow$ | 0.1 | 0.2 | 0.3 | 0.4 | 0.5 | 0.6 | 0.7 | 0.8 | 0.9 |
|---|---|---|---|---|---|---|---|---|---|
| 1 | 0.0987 | 0.1904 | 0.2709 | 0.3393 | 0.3968 | 0.4450 | 0.4855 | 0.5199 | 0.5494 |
| 2 | 0.0497 | 0.0978 | 0.1429 | 0.1842 | 0.2216 | 0.2550 | 0.2840 | 0.3109 | 0.3314 |
| 10 | 0.0010 | 0.0199 | 0.0297 | 0.0394 | 0.0488 | 0.0479 | 0.0668 | 0.0753 | 0.0835 |

From the table, we find that for a fixed $k$, the mode increases with $\theta$ and for a fixed $\lambda$, the mode decreases with $\theta$.

If the random variable $\theta$ has the pdf as given in (1.3), then

$$\begin{aligned}
E(\theta) &= \int_{-\pi}^{\pi} \frac{\theta_1 e^{k\cos\theta_1}}{2\pi I_0(k)}(1+\lambda\sin\theta_1)d\theta_1 \\
&= \int_{-\pi}^{\pi} \frac{\theta_1 e^{k\cos\theta_1}}{2\pi I_0(k)}d\theta_1 + \lambda \int_{-\pi}^{\pi} \frac{\theta_1 e^{k\cos\theta_1}}{2\pi I_0(k)}\sin\theta_1 \, d\theta_1 \\
&= \int_{-\pi}^{\pi} \frac{\theta_1}{2\pi I_0(k)}\left(I_0(k) + 2\sum_{j=1}^{\infty} I_j(k)\cos j\theta_1\right)d\theta_1 + \frac{\lambda}{k}\left(-\theta_1 \frac{e^{k\cos\theta_1}}{2\pi I_0(k)}\right)\Big|_{-\pi}^{\pi} + \frac{\lambda}{k}\int_{-\pi}^{\pi} \frac{e^{k\cos\theta_1}}{2\pi I_0(k)}d\theta_1 \\
&= \frac{\lambda}{2\pi I_0(k)}\left(-\frac{2\pi}{k}e^{-k} + \frac{2\pi}{k}I_0(k)\right) \\
&= -\frac{1}{k}\frac{\lambda}{I_0(k)}\left(e^{-k} - I_0(k)\right) \\
&= \frac{\lambda}{k}\left(1 - \frac{e^{-k}}{I_0(k)}\right)
\end{aligned}$$

We define the $p^{th}$ circular moment $\varphi_p$ of random variable $\theta$ with pdf $f(\theta)$, with $-\pi \leq \theta \leq \pi$ as $\varphi_p = \int_{-\pi}^{\pi} e^{ip\theta} f(\theta)d\theta$.

It is known (see Jammalamadaka and SenGupta (2001)) that if $\varphi_p^*(k)$ is the $p^{th}$ circular moment of von Mises distribution with pdf as given in (1.2) and $\varphi_p(k, \lambda)$ is the $p^{th}$ circular moment of the skewed von Mises distribution with pdf as given in (1.3), then $\varphi_p(k, \lambda)$ and $\varphi_p^*(k)$ are related.

Specifically, we have
$$\varphi_p(k, \lambda) = \varphi_p^*(k) + \frac{i\lambda}{2}\{\varphi_{p-1}^*(k) - \varphi_{p+1}^*(k)\}.$$
Here
$\varphi_p^*(k) = \int_{-\pi}^{\pi} e^{ip\theta} \frac{\theta e^{k\cos\theta}}{2\pi I_0(k)} d\theta$ and $\varphi_p(k, \lambda) = \int_{-\pi}^{\pi} e^{ip\theta} \frac{\theta e^{k\cos\theta}}{2\pi I_0(k)}(1 + \lambda \sin\theta) d\theta.$

## 3. Characterizations

We shall now derive the two characterizations of the skew von Mises distribution. We will need the following two lemmas for the characterization of the skew von Mises distribution by truncated moment.

**Assumption $\mathcal{A}$**
Let $\theta$ be an absolutely continuous random variable with cdf $F(\theta_1)$ and pdf $f(\theta_1)$. Assume that $E(\theta)$ exists and $f(\theta)$ is differentiable. We assume further that $\alpha = \sup\{\theta_1 | F(\theta_1) > 0\}$ and $\beta = \inf\{\theta_1 | F(\theta_1) < 1\}$.

**Lemma 3.1.**
If $E(\theta|\theta \leq \theta_1) = g(\theta_1)\frac{f(\theta_1)}{F(\theta_1)}$, where $g(\theta_1)$ is a continuous differentiable function in $\alpha \leq \theta_1 \leq \beta$, then
$$f(\theta) = c\exp\left(\int \frac{\theta - g'(\theta)d\theta}{g(\theta)}\right),$$
where $c$ is determined by the condition $\int_\alpha^\beta f(\theta_1)d\theta_1 = 1$.
**Proof:**

We know from elementary theory that
$$E(\theta|\theta \leq \theta_1) = \frac{\int_\alpha^{\theta_1} uf(u)\,du}{F(\theta_1)}.$$

Hence, equating the RHS of both sides, we have,
$$g(\theta_1) = \frac{\int_\alpha^{\theta_1} uf(u)\,du}{f(\theta_1)};$$
or equivalently,
$$\int_\alpha^{\theta_1} uf(u)\,du = f(\theta_1)g(\theta_1).$$
Differentiating both sides of the above equation with respect to $\theta_1$, we obtain
$$\theta_1 f(\theta_1) = f(\theta_1)g'(\theta_1) + f'(\theta_1)g(\theta_1).$$
Re-arranging, we get
$$\frac{f'(\theta_1)}{f(\theta_1)} = \frac{\theta_1 - g'(\theta_1)}{g(\theta_1)}.$$

On integrating both sides of the above equation, we obtain
$$f(\theta) = c \cdot \exp\left(\int \frac{\theta - g'(\theta)}{g(\theta)} d\theta\right),$$
where $c$ is determined by the condition $\int_\alpha^\beta f(\theta) d\theta = 1$. This completes the proof.

◀

**Lemma 3.2:** Under the Assumption $\mathcal{A}$, if $E(\theta|\theta \geq \theta_1) = h(\theta_1)\frac{f(\theta_1)}{1-F(\theta_1)}$, where $h(\theta)$ is a continuous differentiable function in $\alpha \leq \theta \leq \beta$, then
$$f(\theta) = c \exp\left(-\int \frac{\theta + h'(\theta)}{h(\theta)} d\theta\right),$$
where $c$ is determined by the condition $\int_\alpha^\beta f(\theta) d\theta = 1$.

**Proof:** We know from elementary theory that
$$E(\theta|\theta \geq \theta_1) = \frac{\int_{\theta_1}^\beta u f(u) du}{1 - F(\theta_1)}.$$

Hence, equating the RHS of both sides, we have,
$$h(\theta_1) = \frac{\int_{\theta_1}^\beta u f(u) du}{f(\theta_1)},$$

or equivalently,
$$h(\theta_1)f(\theta_1) = \int_{\theta_1}^\beta u f(u) du.$$

Differentiating both sides of the above equation with respect to $\theta_1$, we obtain
$$f(\theta_1)h'(\theta_1) + f'(\theta_1)h(\theta_1) = -\theta_1 f(\theta_1)$$
Re-arranging, we get
$$\frac{f'(\theta_1)}{f(\theta_1)} = -\frac{\theta_1 + h'(\theta_1)}{h(\theta_1)}.$$
On integrating both sides of the above equation, we obtain
$$f(\theta) = c \cdot \exp\left(\int -\frac{\theta + h'(\theta)}{h(\theta)} dx\right),$$
where $c$ is determined by the condition $\int_\alpha^\beta f(\theta) d\theta = 1$. This completes the proof.

◀

The following two theorems give the characterizations of von Mises distribution by truncated first moment.

**Theorem 3.1.**
Suppose that the random variable $\theta$ satisfies the conditions given in Assumption $\mathcal{A}$, with pdf $f(\theta)$, cdf $F(\theta)$, $\alpha = -\pi$ and $\beta = \pi$. Then $E(\theta | \theta \le \theta_1) = g(\theta_1)\tau(\theta_1)$, where

$$\tau(\theta_1) = \frac{f(\theta_1)}{F(\theta_1)},$$

$$g(\theta_1) = \frac{2\pi I_0(k) p(\theta_1)}{e^{k\cos\theta_1}(1 + \lambda \sin\theta_1)}$$

and

$$p(\theta_1) = \frac{\theta_1^2 - \pi^2}{4\pi} + \frac{1}{\pi I_0(k)} \sum_{j=1}^{\infty} I_j(k) \left\{ \frac{1}{j} \theta_1 \sin j\theta_1 + \frac{1}{j^2} \cos j\theta_1 - \frac{(-1)^j}{j^2} \right\}$$
$$- \frac{\lambda}{2\pi I_0(k)} \left( \pi e^{-k} + \theta_1 e^{k\cos\theta_1} \right) + \lambda \frac{\theta_1 + \pi}{2k\pi} + \frac{1}{\pi I_0(k)} \sum_{j=1}^{\infty} I_j(k) \frac{1}{j} \sin j\theta_1$$

if and only if $\theta$ has the skew von Mises distribution with pdf $f_\theta(\theta_1)$ as

$$f_\theta(\theta_1) = \frac{e^{k\cos\theta_1}}{2\pi I_0(k)} (1 + \lambda \sin\theta_1),$$

$-\pi \le \theta_1 \le \pi$, $-1 \le \lambda \le 1$ and $k$ is any real number.

**Proof.**
Suppose that

$$f(\theta) = \frac{e^{k\cos\theta}}{2\pi I_0(k)} (1 + \lambda \sin\theta),$$

then

$$g(\theta_1) f(\theta_1) = \int_{-\pi}^{\theta_1} \theta \frac{e^{k\cos\theta}}{2\pi I_0(k)} (1 + \lambda \sin\theta) d\theta$$
$$= \int_{-\pi}^{\theta_1} \frac{\theta e^{k\cos\theta}}{2\pi I_0(k)} d\theta + \lambda \int_{-\pi}^{\theta_1} \frac{\theta e^{k\cos\theta}}{2\pi I_0(k)} \sin\theta \, d\theta$$
$$= \int_{-\pi}^{\theta_1} \frac{\theta}{2\pi I_0(k)} \{I_0(k) + 2\sum_{j=1}^{\infty} I_j(k) \cos j\theta \, d\theta\} + \lambda \left( -\theta \frac{e^{k\cos\theta}}{2\pi I_0(k)} \right)\Big|_{-\pi}^{\theta_1} + \lambda \int_{-\pi}^{\theta_1} \frac{\theta e^{k\cos\theta}}{2\pi I_0(k)} d\theta$$
$$= \frac{\theta_1^2 - \pi^2}{4\pi} + \frac{1}{\pi I_0(k)} \sum_{j=1}^{\infty} I_j(k) \left\{ \frac{1}{j} \theta_1 \sin\theta_1 + \frac{1}{j^2} \cos j\theta_1 - \frac{(-1)^j}{j^2} \right\} - \frac{\lambda}{2\pi I_0(k)} \left( \pi e^{-k} + \theta_1 e^{k\cos\theta_1} \right) +$$
$$\lambda \frac{\theta_1 + \pi}{2k\pi} + \frac{1}{\pi I_0(k)} \sum_{j=1}^{\infty} I_j(k) \frac{1}{j} \sin j\theta_1$$
$$= p(\theta_1), \text{ say}$$

Thus,

$$g(\theta_1) = \frac{2\pi I_0(k)p(\theta_1)}{e^{k\cos\theta_1}(1+\lambda\sin\theta_1)}$$

Next, suppose that
$$g(\theta_1) = \frac{2\pi I_0(k)p(\theta_1)}{e^{k\cos\theta_1}(1+\lambda\sin\theta_1)}.$$

Then
$$g'(\theta_1) = \theta_1 - \frac{2\pi I_0(k)p(\theta_1)}{e^{k\cos\theta_1}(1+\lambda\sin\theta_1)}\left\{-k\sin\theta_1 + \frac{\lambda\cos\theta_1}{1+\lambda\sin\theta_1}\right\}$$
$$= \theta_1 - g(\theta_1)\left\{-k\sin\theta_1 + \frac{\lambda\cos\theta_1}{1+\lambda\sin\theta_1}\right\}.$$

Hence,
$$\frac{\theta_1 - g'(\theta_1)}{g(\theta_1)} = -k\sin\theta_1 + \frac{\lambda\cos\theta_1}{1+\lambda\sin\theta_1}.$$

By Lemma 2.1
$$\frac{f'(\theta_1)}{f(\theta_1)} = -k\sin\theta_1 + \frac{\lambda\cos\theta_1}{1+\lambda\sin\theta_1}$$

On integrating the above equation with respect to $\theta_1$, we obtain
$$f(\theta) = c e^{k\cos\theta}\{1+\lambda\sin\theta\}.$$

Using the condition $\int_{-\pi}^{\pi} f(\theta)d\theta = 1$, we obtain
$$f(\theta) = \frac{e^{k\cos\theta}}{2\pi I_0(k)}\{1+\lambda\sin\theta\}.$$

**Theorem 3.2.**
Suppose that the random variable $\theta$ satisfies the conditions of the Assumption $\mathcal{A}$ with $\alpha = -\pi$ and $\beta = \pi$. Then
$$E(\theta|\theta \geq \theta_1) = \frac{2\pi I_0(k)h(\theta_1)}{e^{k\cos\theta_1}(1+\lambda\sin\theta_1)}r(\theta_1),$$

where
$$r(\theta_1) = \frac{f(\theta_1)}{1-F(\theta_1)}$$

and
$$h(\theta_1) = \frac{2\pi I_0(k)\{E(\theta) - p(\theta_1)\}}{e^{k\cos\theta_1}(1+\lambda\sin\theta_1)}$$

if and only if

$$f(\theta) = \frac{e^{k\cos\theta}}{2\pi I_0(k)}\{1 + \lambda \sin\theta\}.$$

**Proof.**

If the pdf of the random variable $\theta$ is $f(\theta) = \frac{e^{k\cos\theta}}{2\pi I_0(k)}\{1 + \lambda \sin\theta\}$, then

$$f(\theta_1)h(\theta_1) = \int_{\theta_1}^{\pi} \theta f(\theta)d\theta = E(\theta) - p(\theta_1).$$

Thus

$$h(\theta_1) = \frac{2\pi I_0(k)\{E(\theta) - p(\theta_1)\}}{e^{k\cos\theta_1}(1 + \lambda \sin\theta_1)}.$$

Next, suppose that

$$h(\theta_1) = \frac{2\pi I_0(k)\{E(\theta) - p(\theta_1)\}}{e^{k\cos\theta_1}(1 + \lambda \sin\theta_1)},$$

then

$$h'(\theta_1) = -\theta_1 - \frac{2\pi I_0(k)\{E(\theta) - p(\theta_1)\}}{e^{k\cos\theta_1}(1 + \lambda \sin\theta_1)}\left(-k\sin\theta_1 + \frac{\lambda \cos\theta_1}{1 + \lambda \sin\theta_1}\right)$$

$$= -\theta_1 - h(\theta_1)\left(-k\sin\theta_1 + \frac{\lambda \cos\theta_1}{1 + \lambda \sin\theta_1}\right)$$

$$-\frac{\theta_1 + h'(\theta_1)}{h(\theta_1)} = -k\sin\theta_1 + \frac{\lambda \cos\theta_1}{1 + \lambda \sin\theta_1}$$

Thus by Lemma 2.2

$$\frac{f'(\theta_1)}{f(\theta_1)} = -k\sin\theta_1 + \frac{\lambda \cos\theta_1}{1 + \lambda \sin\theta_1}.$$

On integrating the above equation with respect to $\theta_1$, we obtain $f(\theta) = ce^{k\cos\theta}(1 + \lambda \sin\theta)$.

Using the condition $\int_{-\pi}^{\pi} f(\theta)d\theta = 1$, we obtain

$$f(\theta) = \frac{e^{k\cos\theta}}{2\pi I_0(k)}(1 + \lambda \sin\theta).$$

## 4. Concluding Remarks

The characterization of probability distribution plays an important role in probability, statistics and other related fields. Before a particular probability distribution model is applied to fit data in the real world, it is necessary to confirm whether the given probability distribution satisfies the underlying requirements by characterization. A probability distribution can be characterized by various methods. The characterization of probability distribution by truncated moments play an important role in the characterization of distributions by properties in the given data. It is hoped that the findings of the paper will useful for researchers in probability, statistics and other sciences.